\documentclass[lettersize,journal]{IEEEtran}
\usepackage{amsmath,amsfonts}
\usepackage{algorithmic}
\usepackage{algorithm}
\usepackage{array}
\usepackage{textcomp}
\usepackage{stfloats}
\usepackage{url}
\usepackage{verbatim}
\usepackage{graphicx}
\usepackage{cite}
\usepackage{caption}
\usepackage{subcaption}
\usepackage{float}

\usepackage{epsfig}
\usepackage{xcolor}
\usepackage{tikz}
\usepackage{tikz-qtree}
\usepackage{graphicx}

\usepackage{tablefootnote}
\usepackage [para,online,flushleft] {threeparttable}
\usepackage{hhline}
\usepackage{colortbl}
\usepackage{amssymb}

\usepackage{comment}

\usetikzlibrary{trees,positioning,shapes,shadows,arrows}

\definecolor{fashionfuchsia}{rgb}{0.96, 0.0, 0.63}

\begin{document}

\title{A review of Federated Learning in \\ Intrusion Detection Systems for IoT}

\author{Aitor Belenguer, Javier Navaridas and Jose A. Pascual
\IEEEcompsocitemizethanks{\IEEEcompsocthanksitem All authors are with the Department of Computer Architecture and Technology,University of the Basque Country UPV/EHU. Corresponding email: aitor.belenguer@ehu.eus}%
    \thanks{Manuscript received ??; revised ??.}
}

% The paper headers
\markboth{IEEE Internet of Things Journal}{Belenguer \MakeLowercase{\textit{et al.}}: A review of Federated Learning in Intrusion Detection Systems for IoT}

%\IEEEpubid{0000--0000/00\$00.00~\copyright~2022 IEEE}
% Remember, if you use this you must call \IEEEpubidadjcol in the second
% column for its text to clear the IEEEpubid mark.

\maketitle

\begin{abstract}
Intrusion detection systems are evolving into intelligent systems that perform data analysis while searching for anomalies in their environment. The development of deep learning technologies paved the way to build more complex and effective threat detection models. However, training those models may be computationally infeasible in most Internet of Things devices. Current approaches rely on powerful centralized servers that receive data from all their parties -- violating basic privacy constraints and substantially affecting response times and operational costs due to the huge communication overheads. To mitigate these issues, Federated Learning emerged as a promising approach, where different agents collaboratively train a shared model, without exposing training data to others or requiring a compute-intensive centralized infrastructure. This paper focuses on the application of Federated Learning approaches in the field of Intrusion Detection. Both technologies are described in detail and current scientific progress is reviewed and categorized. Finally, the paper highlights the limitations present in recent works and presents some future directions for this technology.
\end{abstract}


%%%%%%%%%%%%%%%%%%%%%%%%%%%%%%%%%%%%%%%%%%%%%%%%%%%%%%%%%%%%%%%%%%%%%%
\section{Introduction}
\IEEEPARstart{I}{n} the era of digitization, the amount of generated and stored data has increased exponentially. The current trend of storing and analyzing any technological interaction, combined with the cheapening of storage devices and infrastructures, has caused an outburst of database instances. In parallel, the number of Internet of Things (IoT) devices is increasing due to the establishment of domestic intelligent gadgets, the spread of smart cities and the rapid advancement of Industry 4.0. Information generated by those devices is highly appreciated by big data conglomerates, which rely on data analysis for Business Intelligence and understanding market trends with the ultimate goal of improving products and services. As a consequence, data has become a highly valuable asset that needs to be protected.

Cybersecurity has become an essential element in order to avoid data leakages, malicious intrusions, service availability denials and so on. However, the area involving information security is uncertain and needs to be constantly readjusted in line with the emergence of new attack paradigms. When cybersecurity firstly appeared, the number of computers was insignificant and they were reserved for professional usage. In those days, fully sensorized smartphones, generating massive network traffic and containing tons of sensitive information, hardly existed. In this context of security preservation, Intrusion Detection Systems (IDS) play an important role by monitoring system activity to proactively detect potential attacks. The evolution of threat detection techniques has evolved in tandem with the development of new Machine Learning (ML) models. The first generation of IDS was rather rudimentary and simply relied on collating system events against manually updated tuples of a signature database. However, these methods were quickly found to have severe limitations, most critically, in terms of flexibility. Primarily, they lacked proactivity in the sense that they were unable to detect new threats that were not in the signature database. Secondly, the period from when an attack was first discovered until new signatures were produced and updated in the IDS was potentially lengthy, leaving the systems vulnerable for long periods of time. 

As a mitigation, second generation IDS started to gradually incorporate some form of intelligence to detect new threats. This way, they were capable of automatically learning attack patterns using basic ML models, e.g., Support Vector Machines (SVM), Random Forests (RF) and so on. The evolution continued with the incorporation of Deep Learning (DL) technologies, which contributed to the advent of more accurate and sophisticated models, e.g., Multilayer Perceptrons (MLP), Recurrent Neural Networks (RNN) and others.

As these systems kept improving in terms of accuracy and new threat detection capabilities, the next natural step is to allow them to share information about newly detected threats so that new attack vectors are promptly recognized by all involved parties and, in turn, global impact is reduced. One possible way of achieving this is the incorporation of centralized learning, in which different parties contribute to the training of a complex model by sending their local data to a centralized computing infrastructure. The whole training process is typically performed in a data center which will then distribute the new model parameters to all involved parties. Nonetheless, performing centralized learning could be infeasible due to traditional information sharing approaches that deal with data in a raw way. That could cause network traffic flow struggle, especially in cases where low resource IoT devices are the main communication agents. Moreover, sharing raw data to third parties is generally discouraged and, indeed, could violate regulations involving data management policies~\cite{law}. Therefore, using collaborative learning algorithms with strict data protection policies is vital to achieve good reliability and scalability, as well as a privacy-friendly infrastructure.

In this context, Federated Learning~\cite{BrendanMcMahan2017} (FL) has emerged as a promising tool to deal with the information exchange of different parties and sensitive data exploitation challenges. FL is an avant-garde ML technique that has gained special interest in edge and IoT computing for its reduced communication cost and privacy preserving features. First, raw data located in the end-devices never leaves these devices -- following an \emph{on-device} policy. Instead, it is used to learn internal models and sharing local model parameters. Then, local parameters from agents are aggregated into a global model following some predefined rules -- e.g., by averaging them as in FedAVG~\cite{BrendanMcMahan2017}. Finally, the consolidated global parameters are sent back to each edge party and the process is iterated until convergence is achieved. Thus, knowledge acquired by collaborating devices is pooled to improve the overall metrics of each local model and obtain improved training scores.

After analyzing the evolution of IDS, we are convinced that FL will conform the backbone of new generation IDS. While FL is a relatively recent technique and its application to IDS technologies is very limited, it is essential to carry out a review of the state-of-the-art to summarize existing knowledge and facilitate future research by highlighting some of limitations of the literature. In particular, this survey discusses the perceived lack of model evaluation standardization and, indeed, proposes a roadmap of good practices to tackle future innovation.

%%%%%%%%%%%%%%%%%%%%%%%%%%%%%%%%%%%%%%%%%%%%%%%%%%%%%%%%%%%%%%%%%%%%%%
\section{Intrusion Detection Systems}
\label{sec:IDS}

IDS can be classified into Host-based (HIDS) and Network-based (NIDS). HIDS are typically computing systems that analyze local system data, application registers, log accesses, system calls and so on in order to detect unwanted applications\cite{B2}. Meanwhile, NIDS focus on network traffic with the aim of finding malicious patterns that target the devices inside a monitored infrastructure~\cite{B3}. Recent research has extensively showed that leveraging ML techniques for intrusion detection is a highly successful methodology to learn complex relationships in the data and, in turn, to build stronger IDS -- both in the context of NIDS and HIDS. 

\subsection{Characteristics of IDS}

Molina et al.~\cite{borja} established a series of basic aspects concerning the establishment and evaluation of IDS:

\begin{enumerate}
    \item \textbf{Input data type} handled, e.g., Raw traffic, data flows and encrypted payloads for NIDS, or application data, memory accesses and kernel operations for HIDS.
    \item \textbf{Location} of the IDS. Either position in the network of a NIDS or the level of the system stack where a HIDS is implemented. 
    \item \textbf{Performance} in terms of accuracy and response time.
    \item \textbf{Adaptability} to rapidly learn new threats.
\end{enumerate}

Moreover, we can classify IDS into three main categories based on the operation of the detector and the interpretation of the attacks. 
\begin{enumerate}
    \item \textbf{Misuse-based detectors (MD)} identify potential threats by relying on attack signatures, patterns or rules which determine nefarious behavior.
    \item \textbf{Anomaly-based detectors (AD)} operate by detecting abnormal patterns that deviate from expected observations or from what is considered as common  behavior~\cite{borj6, borj7, borj8, borj9}.
    \item \textbf{Hybrid detectors} combine misuse and anomaly-based detectors to create more robust systems. These can be further subdivided based on how MD and AD are combined.
    \begin{enumerate}
        \item AD and MD work independently (in parallel) to increase the detection rate~\cite{borj101,borj117}.
        \item MD is concatenated to the AD to filter the anomalies detected by the second one and reduce false positives~\cite{borj56, borj72}. 
        \item AD performs a first filter and redirects what is classified as normal to MD to identify undetected threats. In contrast, what is labeled as an anomaly is redirected to a second AD to reduce false positives~\cite{borj114}.
    \end{enumerate}
\end{enumerate}

Another important aspect of IDS is the learning scheme used to train the models. Based on this, two main categories emerge: \emph{batch learning} and \emph{incremental learning}. 

In batch learning, a data mining algorithm is applied to a set of (labeled or unlabeled) data and predictions are inferred from it. They are not specifically designed for streaming scenarios, and they are usually expensive in terms of time and memory requirements~\cite{borj185}. More complex variants using retraining exist, where stored batch records are used to train a new model and replace the current one~\cite{borj187}. Despite being capable of automatically adjusting the model, new data has to be captured, stored and pre-processed before going through the whole training process again. 

Incremental learning algorithms assume that data is unbounded and arrives in a continuous manner, and also that it may be generated by a time-evolving non-stationary process~\cite{borj186, borj188}. Thus, the model is updated constantly with gradually arriving data samples. Nonetheless, that continuous adaptation makes the model sensitive to noise and, more importantly, to adversarial perturbations that could trick the detector~\cite{borj190}.

\subsection{Datasets for evaluating IDS}
\label{sec:datasets}

\begin{table*}[t]
\centering
\caption{A summary of public datasets available for the evaluation of IDS.}
%\resizebox{.9\textwidth}{!}{
\begin{threeparttable}[t]
\begin{tabular}{l||ccccc||c||c} 

        & Raw      & Payload  & Single flow & Multi flow & Labeling        &        & Year of \\ 
Dataset & captures & features & features    & features   & method \tnote{1} & Domain & capture \\ 
\hline \hline
AWID3~\cite{AWID3}        & \checkmark              &                  &     &     &-                                              &wireless IoT           &2020                  \\
IOT-23~\cite{iot23}     &\checkmark       &\checkmark                  &\checkmark     &\checkmark     &A                                  &IoT       &2020                  \\
TON\_IOT~\cite{ton_iot}    &\checkmark              &                  &\checkmark     &\checkmark     &A               & IoT/IIoT          &2019                  \\
CICIDS-2017~\cite{CICIDS}        &\checkmark              &                  &\checkmark     &     &S                                              & application traffic          &2017                  \\

ISCXTor2016~\cite{tor} &\checkmark              &                  &\checkmark     &     &S                                              & application traffic (raw/Tor)          &     2016             \\
ISCXVPN2016~\cite{vpn} &\checkmark              &                  &\checkmark     &     &S                                              & application traffic (raw/VPN)         &       2016           \\
WSN-DS~\cite{wsn}     &              &                  &\checkmark     &     &S                                              &  wireless IoT         &2016                  \\
AWID2~\cite{AWID2}        & \checkmark              &                  &\checkmark     &     &M                                              &wireless IoT           &2015                  \\
SEA~\cite{sea}        &              &\checkmark                  &     &     &A                                            &  UNIX commands        &2001                  \\
NSL-KDD~\cite{nsl}     &              &\checkmark                  &\checkmark     &\checkmark     &MS                                              & networking                     &1998     \\
\hline
\end{tabular}
\begin{tablenotes}
     \item[1] M: manually, A: automatically, S: scheduled.
   \end{tablenotes}
    \end{threeparttable}%}%
\label{table1}
\end{table*}

Another important aspect in the life-cycle of advanced IDS is the evaluation of their detection capabilities. There are many popular datasets available for IDS evaluation and, indeed, the main datasets discussed in Section~\ref{sec:flids} are gathered together in Table~\ref{table1}. The table uses the following conventions.
\emph{Raw captures} correspond to the availability of the whole captured datagram; usually stored in \textit{pcap} files. \emph{Payload features} are extracted from the application data in the dataframe and processed using natural language processing, regular expressions or similar techniques. \emph{Single Flow Derived Features} (SFD) correspond to a collection of packets sharing any property on the IP and transport layers. SFD features are extracted from the aggregation of a packets flow delimited by a given event (e.g., end of a TCP connection, a timeout and so on). \emph{Multiple Flow Derived Features} (MFD) correspond to the aggregation of information belonging to multiple flow records, containing higher level statistics (e.g., time window delimited flows, last $n$ flows and so on). Finally, dataset labeling could be done \emph{manually} (M) by a skillful professional; \emph{automatically} (A) using a rule repository and a script; or on a \emph{scheduled} (S) way, launching specific attacks in pre-established time windows. It is also possible to merge some of the mentioned methods (MS, AS).

Performing a deeper analysis on the datasets, additional relevant features could be extracted. Regarding the monitored devices, datasets~\cite{iot23, ton_iot} collect data from IoT devices using the Zeek "Bro" network security monitoring tool. Similarly, datasets~\cite{AWID2, AWID3, wsn} capture specific data coming from wireless IoT parties.
The generalized use of DARPA type datasets~\cite{darpa} (i.e., storing and inferring new instances from the information exchange of multiple agents) and its derivatives (e.g.,~\cite{nsl}) are also present in Table~\ref{table1}. Finally, SEA~\cite{sea} is an exception due to its limited scope in detection of masqueraders~\cite{masque} via malicious UNIX command sequences.

However, some of these datasets suffer from the lack of data diversity and volume, some do not cover the variety of known attacks, while others anonymize packet payload data, which cannot reflect the current trends. For that reason, to make data more real-world representative, in datasets~\cite{tor, vpn} the authors collect and characterize traffic from real applications such as web, email, streaming and others over raw/VPN/Tor connections extracting the most representative features using the CICIDS/ISCXFlowMeter tools. A similar approach is carried out by the same authors in the CICIDS-2017 dataset~\cite{CICIDS}, incorporating the most up-to-date common attacks. 

In addition to relying on public datasets for evaluation purposes, many authors rely on custom datasets obtained by capturing real network traffic in either production or testbed systems. This is, indeed, a common practice in IoT specific environments, where network traffic is highly conditioned by the type of hardware, protocol or service provided (e.g., eHealth networks, industrial control systems, SCADA~\cite{turni} data collection).

%%%%%%%%%%%%%%%%%%%%%%%%%%%%%%%%%%%%%%%%%%%%%%%%%%%%%%%%%%%%%%%%%%%%%%
\section{Federated Learning}
\label{sec:fl}

With the aim of achieving greater understanding of state-of-the-art FL-IDS, this section provides useful background information on FL. 
Although FL is the main focus of this paper, we also address three additional learning paradigms which are typically used as baselines for benchmark comparisons within IoT/Edge infrastructures.

\begin{enumerate}
    \item \textbf{Self learning (SL)}:  Neither data nor parameters leave the device;  training is performed individually by edge devices. SL can be used as a baseline for learning ability  when no information is shared.
    \item \textbf{Centralized learning (CNL)}:  Data is sent from different parties to a centralized computing infrastructure, which is in charge of performing the training with all the received data. CNL is used as a yardstick of the learning ability when the models are built using all available data.
    \item \textbf{Collaborative learning (CL)}: Wraps up custom variants of distributed learning (including FL) where involved agents benefit from training a model jointly. Paul Vanhaesebrouck et al.~\cite{lille} presented a fully decentralized collaborative learning system, where the locally learned parameters are spread and averaged without being under the orchestration of a centralized authority -- in a P2P network.
\end{enumerate}

According to Chaoyang He et al.~\cite{fedml}, the main limitations of FL when compared with CNL are concerning statistical heterogeneity, system constraints and trustworthiness. Those challenges have been addressed using different approaches in the literature. For instance, statistical heterogeneity has been tackled by distributed optimization methods such as Adaptive Federated Optimizer~\cite{adaptive}, FedNova~\cite{fnova}, FedProx~\cite{fprox} and FedMA~\cite{fma}. System constraints, such as communication overheads or high training computation costs~\cite{fml6, fml7, fml8, fml9, fml10, fml11, fml12} are mitigated using gradient sparsification~\cite{D30} and quantization techniques. Finally, to tackle trustworthiness issues, Differential Privacy (DP) and secure multiparty computation (SMPC) privacy mechanisms have been proposed~\cite{fml25, fml26, fml27, fml28, fml29, fml30, fml31, fml32, fml33}. Similarly, new defense techniques to make FL robust against adversarial attacks have been proposed as well~\cite{fml13, fml14, fml15, fml16, fml17, fml18, fml19, fml20, fml21, fml22}.

\subsection{Federated Learning Systems}
\label{sec:fls}
By definition, FL enables multiple parties to jointly train a ML model without exchanging local data. It involves distributed systems, ML and privacy research areas~\cite{surv1,surv2}, and, since the pioneer FedAVG~\cite{BrendanMcMahan2017} approach, many new Federated Learning Systems (FLS) have emerged. A general taxonomy describing the difference of those FLS is presented in~\cite{surv2} and replicated in Figure~\ref{fig:FLS}. This classification is multidimensional and includes the most important aspects of FL architectures including data partitioning, learning model, privacy, communication characteristics and so on.

\tikzset{
  basic/.style  = {draw, text width=7cm, drop shadow, font=\sffamily, rectangle},
  root/.style   = {basic, rounded corners=0pt, thin, align=center, fill=white},
  l1/.style={sibling distance=10cm, level distance=10em},
  l2/.style = {basic, rounded corners=4pt,sibling distance=10cm, thin,align=center, fill=white, text width=3.2cm},
  l3/.style = {basic, thin, align=center, fill=white, text width=2.6cm}
}
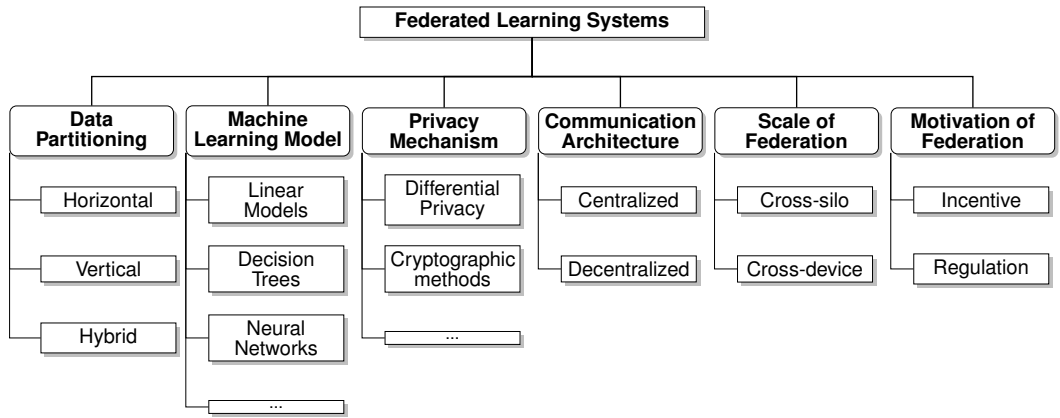
\begin{figure*}[t]
    \centering
\resizebox{400pt}{!}{
\begin{tikzpicture}[
    edge from parent/.style={solid,black,thick,sloped,draw},
    edge from parent path={(\tikzparentnode.south) -- (\tikzchildnode.north)},
    >=latex, node distance=1.45cm, edge from parent fork down]

\node[root] {\textbf{\large Federated Learning Systems}}
    child [sibling distance=37mm, level distance=65pt] {node[l2] (b1) {\textbf{\large Data Partitioning}}}
    child [sibling distance=37mm, level distance=65pt] {node[l2] (b2) {\textbf{\large Machine Learning Model}}}
    child [sibling distance=37mm, level distance=65pt] {node[l2] (b3) {\textbf{\large Privacy Mechanism}}}
    child [sibling distance=37mm, level distance=65pt] {node[l2] (b4) {\textbf{\large Communication Architecture}}}
    child [sibling distance=37mm, level distance=65pt] {node[l2] (b5) {\textbf{\large Scale of Federation}}}
    child [sibling distance=37mm, level distance=65pt] {node[l2] (b6) {\textbf{\large Motivation of Federation}}}
    ;

\begin{scope}[every node/.style={l3}]
\node [below of = b1, xshift=10pt,] (b11) {\large Horizontal};
\node [below of = b11] (b12) {\large Vertical};
\node [below of = b12] (b13) {\large Hybrid};

\node [below of = b2, xshift=5pt] (b21) {\large Linear Models};
\node [below of = b21] (b22) {\large Decision Trees};
\node [below of = b22] (b23) {\large Neural Networks};
\node [below of = b23] (b24) {...};

\node [below of = b3, xshift=5pt] (b31) {\large Differential Privacy};
\node [below of = b31] (b32) {\large Cryptographic methods};
\node [below of = b32] (b33) {...};

\node [below of = b4, xshift=5pt] (b41) {\large Centralized} ;
\node [below of = b41] (b42) {\large Decentralized};

\node [below of = b5, xshift=5pt] (b51) {\large Cross-silo} ;
\node [below of = b51] (b52) {\large Cross-device};

\node [below of = b6, xshift=5pt] (b61) {\large Incentive} ;
\node [below of = b61] (b62) {\large Regulation};
\end{scope}

% draw lines
\foreach \value in {1,...,3}
  \draw[] (b1.195) |- (b1\value.west);

\foreach \value in {1,...,4}
  \draw[] (b2.195) |- (b2\value.west);

\foreach \value in {1,...,3}
  \draw[] (b3.195) |- (b3\value.west);
  
\foreach \value in {1,...,2}
  \draw[] (b4.195) |- (b4\value.west);
  
\foreach \value in {1,...,2}
  \draw[] (b5.195) |- (b5\value.west);
  
\foreach \value in {1,...,2}
  \draw[] (b6.195) |- (b6\value.west);

\end{tikzpicture}
}
    
\caption{A big picture classification of existing Federated Learning Systems presented by Q.Li et Al.~\cite{surv2}.}
\label{fig:FLS}
\end{figure*}

FLS can handle a large variety of \textbf{Machine Learning Models}. Provided that learning convergence is ensured, the possibilities for the learning model have no bounds, from simple ones, such as linear models and decision trees, to more complex approaches, such as (deep) neural networks and others. The most common categories of learning models employed for FL, together with some pros and cons, are described below:

\begin{enumerate}
    \item \textbf{Linear models}: Simple and computationally efficient models in which the output is easily interpreted. However, they assume linearity and independence among the variables, which is rarely true in real world scenarios.
    \item \textbf{Decision trees}: Easy to understand and interpret, while making no assumptions about the shape of the data (non-parametric model). However, they could be prone to overfitting and their learning process could be exceedingly resource consuming.
    \item \textbf{Neural networks}: Modeling approach to capture efficiently non-linearity in the data, in which the inference process is very fast. However, as they depend a lot on training data, the problem of overfitting and generalization appears often. Also, the learning process is compute-intensive, making the utilization of specialized computing hardware (GPUs, TPUs and others) essential.
\end{enumerate}

To avoid inversion or membership inference attacks~\cite{S56,S167}, the parameters of the learned model should not share sensitive information. In order to do so, different \textbf{privacy mechanisms} are presented:
\begin{enumerate}
    \item \textbf{Cryptographic methods}: Message encryption before sending it through the network; homomorphic encryption~\cite{S15} and secure multi-party computation (SMPC)~\cite{S165}.
    \item \textbf{k-Anonymity}: Refers to the property of having at least k-1 instances belonging to the same set of identical instances -- achieved after modifying a dataset by performing feature grouping, omission and so on~\cite{S50}.
    \item \textbf{Differential privacy}: Guarantees that a single record does not provide much information about the output of a given function. A usual mechanism to deal with this issue is (Laplacian) noise addition~\cite{S48,S49}.
\end{enumerate}

Depending on the information exchange paradigm, centralized and decentralized \textbf{communication architectures} emerge. 
\begin{enumerate}
    \item \textbf{Centralized}: Model aggregation is always performed in a global server and results are sent back to local parties. The communication is often asymmetric and it can be synchronous~\cite{BrendanMcMahan2017} or asynchronous~\cite{fls171, fls204}.
    \item \textbf{Decentralized}: Model parameters are shared in a distributed fashion, typically through point-to-point communication with a subset of the peers. It involves fairness and communication overhead challenges and, indeed, generates new issues concerning consensus preservation and more complex convergence guarantees. Three decentralized designs are discussed: P2P~\cite{S104,S217}, graph~\cite{S128} and blockchain~\cite{S191}.
\end{enumerate}

The \textbf{Scale of Federation} is framed by the features of the agents taking part in the learning process.
\begin{enumerate}
    \item \textbf{Cross-silo}: Small number of parties with high computational resources, usually corporations and data centers.
    \item \textbf{Cross-device}: High number of parties with limited computational resources, usually smartphones and IoT devices.
\end{enumerate}

Three types of \textbf{data partitioning} schemes exist, depending on how data is split among the system participants. Furthermore, each scheme has its own independent and identically distributed (IID) and non-IID variant depending on how the information is distributed across agents~\cite{IID}. 
\begin{enumerate}
    \item \textbf{Horizontal}: Datasets of different parties have the same feature space, but little intersection on the sample space (typical in non-IID type environments). 
    \item \textbf{Vertical}: Agents have similar sample spaces, but they differ in the feature space.
    \item \textbf{Hybrid:} Both the feature and sample spaces are different in each agent.
\end{enumerate}

Finally, \textbf{Motivation of Federation} justifies why individual parties enroll in a given FLS.
\begin{enumerate}
    \item \textbf{Incentives}: Mutual benefits from learning accurate models without the need to release valuable data (e.g., companies training product quality evaluation systems).
    \item \textbf{Regulations}: In many domains, complying with laws, policies, standards and certifications prevent institutions from sharing their data. Therefore, sharing models via FL (or CL) is the only alternative. For instance, patient data in hospitals is heavily regulated, but medical research benefits greatly from cross-site, international collaborations.  
\end{enumerate}

\subsection{Federated Learning Frameworks}

There exist many available frameworks and libraries which can be used to develop FL applications, as thoroughly discussed in~\cite{fedml}. Frameworks can be categorized based on their main objectives: \emph{Simulation-oriented} libraries provide multiple development and benchmark tools, placing the emphasis on extensibility (adding new functionalities) and evaluation purposes (e.g., using emulation and virtual devices). In contrast, \emph{Production-oriented} libraries offer enterprise level solutions, giving support to various FL scenarios, by focusing on usability and productivity (i.e., facilitating system deployment).

Examples of simulation oriented libraries are TensorFlow-Federated (TFF)~\cite{fml39}, PySyft~\cite{fml28}, LEAF~\cite{fml40} and FedML~\cite{fedml}. Conversely, instances of production oriented libraries are FATE~\cite{fml42} and PaddleFL~\cite{fml43}. Among the APIs mentioned, FedML and PySyft pave the way for the creation of adaptable systems, providing FLS topology and message exchange customization. Nonetheless, regarding the disposal of the parties, all the aforementioned libraries support vanilla FL centralized algorithms (e.g., FedAVG, FedProx).
In the same way, FedML, FATE and PaddleFL exclusively incorporate vertical data partitioning. Bearing all this in mind, FedML seems to be the most complete research-oriented library, in terms of supporting multiple FL setups. Additionally, it simplifies codification with a modular worker/client-oriented architecture.

%%%%%%%%%%%%%%%%%%%%%%%%%%%%%%%%%%%%%%%%%%%%%%%%%%%%%%%%%%%%%%%%%%%%%%
\section{Employing Federated Learning Systems for Intrusion Detection}
\label{sec:flids}

The combination of both previously explained technologies (IDS in Section~\ref{sec:IDS} and FL in Section~\ref{sec:fl}) has become a hot topic of research. Considering that the overwhelming majority of IDS rely on DL models, we introduce a taxonomy based on the DL variants employed by the FL-IDS literature illustrated in Figure~\ref{fig:tax}.

The proposed classification is performed taking into account the DL model architecture used on each edge device. Since vanilla FL is the \textit{de facto} implementation choice and that many disjointed custom variants of it exist, it is not possible to perform a tree structure taxonomy by type of FL algorithm used. Hence, existing FL-IDS are split into two major groups depending on the NN type; Recurrent Neural Networks (RNN)~\cite{RNN} and Multilayer Perceptrons (MLP)~\cite{MLP}. Each group is respectively divided into two subgroups. The RNN models are  divided based on the neuron subtype into Long Short-Term Memory (LSTM)~\cite{LSTM} and Gated Recurrent Units (GRU)~\cite{GRU}. In contrast, MLP models are divided by the NN architecture. In particular, Autoencoders (AE)~\cite{AE} is considered an important subclass because it is commonly employed in the literature.

%%%%%%%%%%%%%%%%%%%
\tikzset{
  basic/.style  = {draw, text width=2cm, drop shadow, font=\sffamily, rectangle},
  root/.style   = {basic, rounded corners=0pt, thin, align=center, fill=white},
  level 1/.style={sibling distance=10em, level distance=5em},
  level-2/.style = {basic, rounded corners=4pt,sibling distance=12em, thin,align=center, fill=white, text width=3cm},
  level-3/.style = {basic, rounded corners=8pt,sibling distance=6em, thin,align=center, fill=white, text width=3cm},
  level-4/.style = {basic, thin, align=center, fill=white, text width=2.2cm}
}
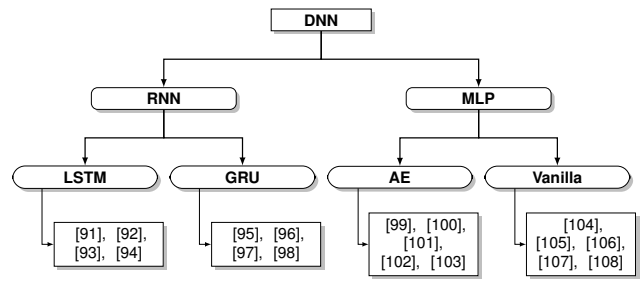
\begin{figure}[t]
    \centering
\resizebox{240pt}{!}{
\begin{tikzpicture}[
    edge from parent/.style={->,solid,black,thick,sloped,draw},
    edge from parent path={(\tikzparentnode.south) -- (\tikzchildnode.north)},
    >=latex, node distance=1.5cm, edge from parent fork down]

\node[root] {\textbf{DNN}}
    child {node[level-2] (c1) {\textbf{RNN}}
        child {node[level-3] (c11) {\textbf{LSTM}}}
        child {node[level-3] (c12) {\textbf{GRU}}}}
    child {node {} edge from parent[draw=none]}
    child {node[level-2] (c2) {\textbf{MLP}}
        child {node[level-3] (c22) {\textbf{AE}}}
        child {node[level-3] (c23) {\textbf{Vanilla}}}};

\begin{scope}[every node/.style={level-4}]
\node [below of = c11, xshift=15pt,] (c111) {\cite{Liu2021}, ~\cite{Huong2021}, ~\cite{Zhao2020}, ~\cite{Lin2020}};

\node [below of = c12, xshift=15pt] (c121) {\cite{Li2021}, ~\cite{Nguyen2019}, ~\cite{Mothukuri2021}, ~\cite{Chen2020}};

\node [below of = c22, xshift=15pt] (c221) {\cite{Qin2021}, ~\cite{Preuveneers2018}, ~\cite{Cholakoska2021}, ~\cite{Cetin2019}, ~\cite{Tian2021}};

\node [below of = c23, xshift=15pt] (c231) {\cite{Weinger2020}, ~\cite{Khramtsova2020}, ~\cite{Al-AthbaAl-Marri2020}, ~\cite{Rahman2020}, ~\cite{Zhao2019}} ;
\end{scope}

% draw lines
\foreach \value in {1}
  \draw[->] (c11.195) |- (c11\value.west);

\foreach \value in {1}
  \draw[->] (c12.195) |- (c12\value.west);

\foreach \value in {1}
  \draw[->] (c22.195) |- (c22\value.west);
  
\foreach \value in {1}
  \draw[->] (c23.195) |- (c23\value.west);

\end{tikzpicture}
}
%%%%%%%%%%%%%%%%%%%
    
\caption{Existing Deep Learning Federated Intrusion Detection Systems by model architecture.}
\label{fig:tax}
\end{figure}

\subsection{Long Short-Term Memory}
\label{lab:lstm}
The utilization of LSTM NNs is interesting due to their ability to process data sequences. If network traffic flow is considered as a time series, it becomes a suitable input for a NN using LSTM neurons. The following lines summarize several FL-IDS developed with this RNN subtype.

Although Liu et al.~\cite{Liu2021} do not explicitly focus on ID, they introduce a novel anomaly detection technique that could be extrapolated to the cybersecurity field. The work presents an FL framework to enable IoT devices to collaboratively train a given model, as well as an Attention Mechanism-based Convolutional Neural Network Long Short-Term Memory (AMCNN-LSTM) architecture to perform anomaly detection. Additionally, it proposes a gradient compression mechanism to improve communication efficiency. The model achieves overall good detection accuracy, over 92\%, in the four real-world datasets tested: power demand, space shuttle, ECG and engine. Moreover, it obtains the lowest Root Mean Squared Error (RMSE) compared to other state-of-the-art systems: CNN-LSTM~\cite{D42}, LSTM~\cite{D41}, GRUs~\cite{D43}, Stacked Autoencoders (SAE)~\cite{D44} and Support Vector Machines (SVM)~\cite{D45}. The system is inspired in previous Deep Anomaly Detection (DAD) approaches~\cite{DAD}, specifically in those using LSTM~\cite{D12, D13, D14} and CNN DeepAnT~\cite{D15} techniques. First, they introduce an attention mechanism system to improve the focus on important features, which are extracted from time series data by a CNN unit. Secondly, they use an LSTM module to perform anomaly detection. Then, only the gradients with larger absolute values are selected for model aggregation. They found that 99.9\% of the gradient exchange is redundant, so they leverage gradient sparsification~\cite{D30} as the main gradient compression method~\cite{D34}. Momentum correction and local gradient clipping~\cite{D30} are used to mitigate the model divergence caused by high sparsification values. As a result, network traffic is considerably reduced.

Huong et al.~\cite{Huong2021} proposed a cyberattack anomaly detection system for Industrial Internet of Things (IIoT) systems. In order to detect anomalies in time series data, they use a model architecture composed of an VAE-Encoder, an LSTM unit and an VAE-Decoder. Moreover, an optimized threshold using Kernel Quantile Estimator (KQE) is learned to achieve high anomaly discrimination accuracy. The main experimentation is performed using data from a Gas Pipeline Factory, acquired from SCADA systems~\cite{turni}. The results show that the approach gets a 97.9\% F1-score. Furthermore, a substantial bandwidth reduction of 35\% is obtained when compared with a centralized learning architecture.

Zhao et al.~\cite{Zhao2020} proposed an Intelligent Intrusion Detection system to detect attacks in UNIX command sequences. A tokenizer is used to convert command line text into vectorized shell command blocks. Then, command blocks are used as the input data for the model. The SEA command record dataset is used to deliver the experimentation. Moreover, two IDS approaches are proposed using centralized learning before implementing the FL model; LSTM and CNN models. The LSTM model outperformed the CNN in all metrics, so it was the selected architecture for the FLS. The proposed FL-LSTM system achieved a 99.21\% F1-score.

An FL malware classification system is proposed by Lin et al.~\cite{Lin2020}, which is not intended to be an IDS. However, it combines FL with DL architectures such as LSTM and SVM. The dataset used for the experimentation is provided by Virustotal. The system gets a 91.67\% classification accuracy in the best scenario (LSTM).

\subsection{Gated Recurrent Units}
Similarly to LSTMs, GRU neurons are a good candidate for processing time series. In contrast to LSTMs, they do not contain an internal memory. However, their simpler architecture makes the learning process lighter which, in turn, renders it suitable for low resource IoT scenarios. Previous research relying on GRU models for FL-IDS is covered in this section.

DeepFed~\cite{Li2021} is a federated deep learning scheme that agglutinates a DL architecture and an FL framework with Paillier cryptosystem~\cite{pallier} privacy mechanism to perform ID in cyber-physical systems (CPS). The system is composed of a Trust Authority, a Cloud Server and several Industrial Agents. The Trust Authority is in charge of managing the public/private keys to be used in secure communications, whereas the Cloud Server is responsible for building and sharing the learned ID model via FedAVG. This way, each Industrial Agent learns its own ID (local) model, based on the data collected from a series of monitored CPS devices and updates its parameters by recurrently interacting with the Cloud Server. The experiments carried out show that the lowest F1-score obtained on a real, time series operating, industrial CPS dataset is 97.34\%. The proposed ID model is based on a CNN-GRU architecture, composed of: (1) a CNN module of 3 convolutional blocks, each of them with a batch normalization and a max-pooling layer; (2) a GRU module of two identical GRU layers; (3) a concatenation between the previous modules, linked to an MLP block of two fully connected layers plus a dropout layer; (4) a softmax output layer.

Dïot~\cite{Nguyen2019} presents an autonomous self-learning distributed system for detecting compromised IoT devices. It achieves an accuracy of 95.6\% with no false positives in real-world smart home deployments. The architecture of the system is based on a Security Gateway (local) and an IoT Security Service (global). The Security Gateway identifies all the devices connected to its network and hosts the anomaly detection component. That component monitors and judges the behavior of each device, based on the locally learned model, which is aggregated by the IoT Security Service. Therefore, the IoT Security Service maintains a global repository of device-type-specific anomaly detection models; expecting that devices of the same type produce similar behaviors. The project is focused on training GRU based models to identify Mirai~\cite{mirai} infected devices by training the models with custom generated network traffic. A dataset is generated by collecting network activity in normal, deployment and infected stages. Although the system is only tested under Mirai attack detection, the authors claim that it is  general enough so as to perform similarly well with other malware botnets such as Perisai and Hajime~\cite{hajime}. However, it has a series of limitations; (1) the evolution of the behavior of a specific device could enter into conflict with the learned model (e.g., after a firmware update); (2) an infected device could mimic legitimate communication patterns to try to remain undetected; (3) an infected device could send specific packets through the network; poisoning the learned model and preventing malicious activities from being detected~\cite{carlini2017}.

Mothukuri et al.~\cite{Mothukuri2021} propose a decentralized FL approach with an ensembler to perform intrusion detection in IoT networks using decentralized on-device data. LSTM and GRUs are used over the Modbus network dataset~\cite{modbus}, achieving an average detection accuracy of 90.25\%. Using PySyft DL framework~\cite{ryffel2018generic}, a series of Virtual Instances (VI), replicating IoT devices and a central server are created. Global network data is captured in the form of \textit{pcap} files and converted into CSV files to perform feature elimination and, then, be equally split into chunks among the VIs. The training is performed asynchronously with the available IoT instances. 4 Different GRU models are built, with different hidden layer sizes, number of layers and dropout rates. The architecture, achieving best results (in the FL scenario), is composed of 2 hidden layers of 256 neurons with a dropout rate of 0.01. Moreover, different window sizes are defined to measure training performance and time against non-FL. Finally, Ensemble Learning~\cite{Dietterichl2002-DIEEL} in the form of Random Forest decision tree classifier~\cite{rf} is used to combine the outputs of ML models and achieve higher accuracy rates.

FedAGRU~\cite{Chen2020} proposes an IDS in Wireless Edge Networks (WENs), combining GRU and SVM models under a custom FL algorithm. FedAGRU uses Attention Mechanism to calculate the importance of the uploaded model parameters. This is done with the aim of both measuring the performance improvement of the global model and sorting the clients according to their importance. This way, in bandwidth scarcity scenarios, important clients will preferentially upload their parameters to the server. Moreover, if a local model has an importance below a pre-established threshold and exceeds a pre-defined counter without providing relevant data, only global model parameters will be accepted for local updating; following the policies proposed in Communication-Mitigated FL (CMFL)~\cite{CMFL}. As a result, the communication overhead will be considerably reduced and model convergence ensured. Additionally, FedAGRU shows a high robustness against poisoning attacks by the suppression of the poisoned samples due to their low impact on the global classification performance. KDD-CUP99~\cite{KDDCUP}, CICIDS-2017~\cite{CICIDS} and WSN-DS~\cite{wsn} datasets are used to evaluate the versatility of the proposed system in different threat detection environments. An F1-score of 97.12\% is achieved in the Non-IID scenario of the WSN-DS dataset. On top of that, the system showed a 70\% reduction in terms of per-round communication compared to vanilla FedAVG, which is translated into faster convergence and higher scalability.

\subsection{Autoencoders}
AEs are the most common FL-IDS architecture~\cite{Qin2021, Tian2021, Preuveneers2018, Cholakoska2021,Cetin2019} to perform ID via anomaly detection due to their input reconstruction abilities. Once the usual network traffic patterns are learned, anomalies are translated into high reconstruction loss instances. The following lines cover the main approaches of this architecture using MLP neurons.

Qin et al.~\cite{Qin2021} face the challenge of performing real-time high-dimensional time series analysis to create an anomaly-based ID for resource-limited embedded systems. On-device sequential learning neural network ONLAD~\cite{tsukada2020neural} is created by applying Online Sequential Extreme Learning Machine (OS-ELM)~\cite{4012031} to an Autoencoder~\cite{doi:10.1126/science.1127647}, with the objective of performing threat detection. Moreover, a greedy feature selection algorithm is employed to deal with data dimensionality issues and obtain better detection results. Furthermore, FedAVG is used to group the devices according to their target attack types so that possible accuracy reductions caused by the aggregation of all the models are avoided. As a consequence of feature selection, only the data owners with the same feature space will establish a federated global model. NSL-KDD~\cite{nsl} dataset is used to evaluate the performance of the system. As previously mentioned, OS-ELM is a sequential implementation of batch learning applied to an autoencoder. Its architecture is composed of variable input and output size layers (depending on the feature selection approach) with a fixed size hidden layer of 64 neurons. The maximum detection accuracy was 70.4\%, whereas the overall accuracy improvement compared to the worst case (without feature selection) was measured to be 25.7\%.

Tian et al.~\cite{Tian2021} propose a Delay Compensated Adam (DC-Adam) approach~\cite{adam} to overcome gradient delay, inconsistency issues, caused by performing anomaly detection in IoT and CPS heterogeneous devices. Additionally, they present a pre-shared data training strategy to avoid model divergence in non-IID data scenarios. The system is tested over the MNIST~\cite{MNIST}, CICIDS-2017~\cite{CICIDS} and IoT-23~\cite{iot23} datasets, obtaining 91.65\%, 92.92\% and 90.15\% F1-score results respectively. One of its main peculiarities is the use of a warm-up strategy for Non-IID data. This is based on randomly selecting a set of $k$ nodes that will share a parameterized fraction of their dataset (5\%-10\%) with the server to pre-train the global model. Although the privacy preservation rule of FL is violated, its impact is greatly mitigated by sharing a small proportion of the datasets. Furthermore, the asynchronous model training with delayed gradient compensation, aggregates information without waiting for all the learning parties to upload their local parameters (i.e., model aggregation is performed as soon as partial updates are received). Therefore, arriving gradients could be obsolete (delayed) for the server current model state. This inconsistency is attenuated using the Taylor Expression formula~\cite{zheng2020asynchronous}, which reduces oscillation of the loss function while minimizing it. To perform anomaly detection, a Denoising Autoencoder (DAE) is proposed with input and output layers of size $n$ and three hidden layers of sizes $\frac{n}{2}, \frac{n}{4}$ and $\frac{n}{2}$. The Mean Squared Error (MSE) between the input and the reconstructed output is measured and compared with a predefined threshold to classify the input as anomalous or normal traffic.

A permissioned blockchain-based~\cite{BC} FLS is proposed by Preuveneers et al.~\cite{Preuveneers2018}, where model updates are chained on the distribution ledger with the objective of supporting decentralized ML model auditing. Although the training process is transparent, the introduction of blockchain transactions burdens global performance by between 5\% and 15\%. The system paves the way to smart contracts audition via Hyperledger Fabric blockchain~\cite{hyperledg}. The proposed approach makes the FLS more robust against poisoning attacks. They use a vanilla autoencoder model architecture and test it over the CICIDS-2017 dataset, obtaining 97\% accuracy in both training and validation phases.

Cetin et al.~\cite{Cetin2019} propose a wireless IDS which uses a Stacked Autoencoder (SAE) and FedAVG to perform ID on the AWID dataset~\cite{AWID2}. Moreover, autoencoder based FL-IDS could be extended to other areas of edge computing. For instance, Cholakoska et al.~\cite{Cholakoska2021} leverage FL to deal with anomalies in eHealth networks. Due to the sensitive nature of medical data, employing Differential Privacy gains special attention in this work. 

\subsection{Vanilla \& others}

\begin{figure*}[t]
    \centering
    \begin{subfigure}{\columnwidth}%[ROC space (TPR vs FPR).]{}
        \centering
        \includegraphics[width = .9\columnwidth]{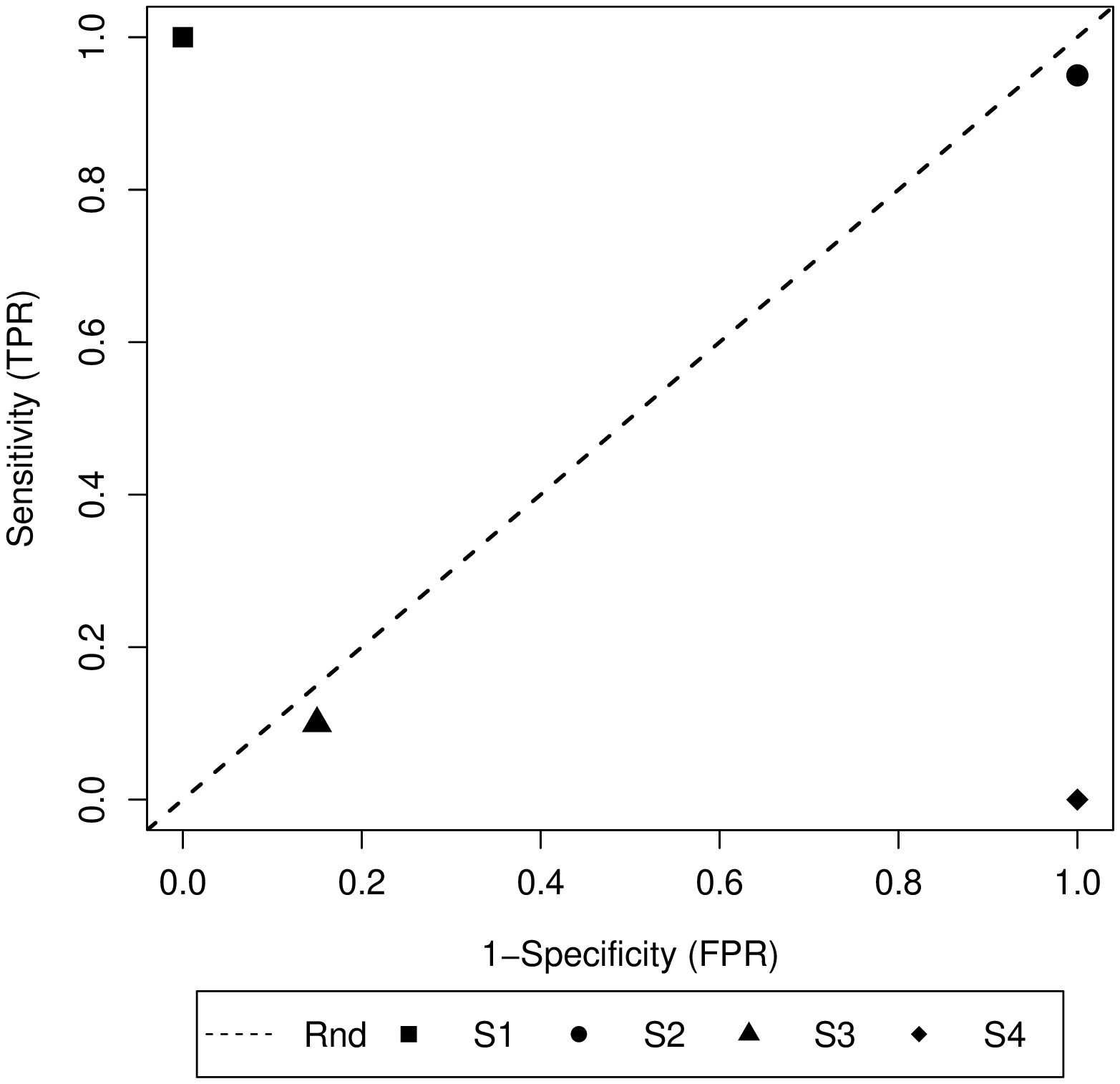}
         \caption{ROC space (TPR vs FPR).}
   \end{subfigure}
    \begin{subfigure}{\columnwidth}%[PR space (PPV vs TPR).{}
        \centering
        \includegraphics[width = .9\columnwidth]{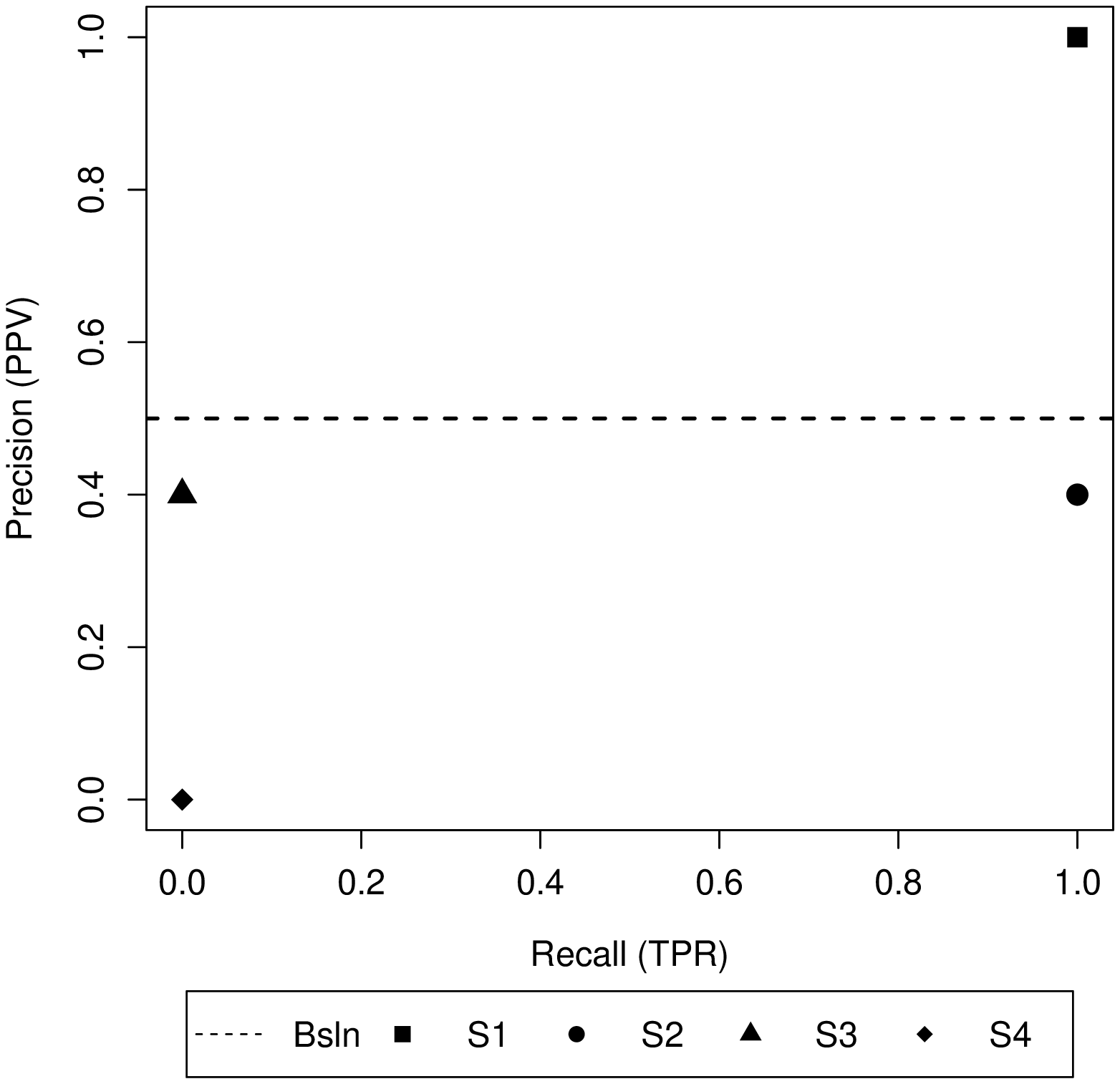}
         \caption{PR space (PPV vs TPR).}
    \end{subfigure}
    \caption{Example of the 2D graphical representation of the predictive performance metrics with the four illustrative scenarios.}
    \label{fig:roc_pr}
\end{figure*}

FL-IDS using custom MLP NN architectures are presented in~\cite{Al-AthbaAl-Marri2020, Zhao2019, Rahman2020, Weinger2020}. Beyond the mentioned architectures, the following lines review pioneer preprocessing and learning variants in FL-IDS approaches.

Al-Marri et al.~\cite{Al-AthbaAl-Marri2020} merge the advantages of FL and mimic learning~\cite{mimic} to create a distributed IDS that achieves a 98.11\% detection accuracy for the NSL-KDD~\cite{nsl} dataset. Following the mimic learning paradigm, a teacher and a student model are learned per device. The teacher models are created using the private data of the users. Then, the teacher models are employed to label the unlabeled public dataset of each user, which are, in turn, used to train the student models. Finally, FedAVG is performed with the learned models in a central server. There are two types of methods depending on which trained model the learned global parameters are added to (teacher or student): Federated Teacher Mimic Learning (FTML) and Federated Student Mimic Learning (FSML). Feedforward Multilayer Perceptrons (MLPs) are used as the NN backbone of the IDS. The architecture is composed of two hidden layers with 256 neural units each (using ReLu activation); followed by a dropout rate of 0.4 and a final softmax layer. ADAM optimizer is used to train the model, with a batch size of 128 and a learning rate of 0.001. Additional preprocessing techniques such as one-hot encoding and feature elimination are employed as well. Both FTML and FSML produced nearly identical outcomes.

A multi-task (MT) network anomaly detection system ``MT-DNN-FL'' is proposed by Zhao et al.~\cite{Zhao2019} to perform anomaly detection, VPN or Tor traffic recognition and traffic classification tasks simultaneously. The experimentation is conducted over CICIDS-2017, ISCXVPN-2016 and ISCXTor-2016 datasets. MT learning aims to concurrently learn multiple related tasks given a common input~\cite{MOL}. Its goal is to train a model that maps an input space with multiple output spaces, where each output presents one objective task. The proposed network architecture is composed of input, shared and task layers: The input layer connects the external inputs and the shared layers. The shared layers extract common abstract features from the received parameters. The task layers are subnetworks with specific tasks connected to the shared layers. Then, FedAVG is used to aggregate model parameters. The performance evaluation is done in two scenarios: In scenario A, CICIDS-2017 and ISCXVPN-2016 datasets are used, achieving a precision over 94.77\% and a recall over 94.70\% in the three mentioned tasks. In scenario B, CICIDS-2017 and ISCXTor-2016 datasets are used; achieving a precision over 94.19\% and a recall over 96.31\% in the same three tasks.

Rahman et al.~\cite{Rahman2020} made a comparison among FL, SL and CNL in IoT intrusion detection duty. They perform one-hot encoding and Min-Max normalization over the NSL-KDD dataset, in order to feed a 122x288x2 NN (used for threat detection) with a clean input. The experiments, performed over Raspberry Pi devices, showed that FL outperforms SL and, in fact, obtains results comparable to those of CNL.

With the aim of enhancing IoT anomaly detection performance, an interesting preprocessing technique is proposed by Weinger et al.~\cite{Weinger2020}. They show how data augmentation could attenuate vanilla FL performance degradation and improve overall classification results in IID scenarios. Data augmentation is performed by random sampling, using SMOTE~\cite{smote} and ADASYN~\cite{adasyn} techniques. Performance evaluation is done by splitting the Modbus dataset among 35 nodes in a homogeneous (equal sizes) and heterogeneous way. The best F1-score results obtained respectively after 100 rounds of FedAVG are approximately 78\% and 70\% by random sampling. 
Interestingly, data augmentation is very effective in improving convergence during the first few rounds of aggregation, but stops being profitable after 30-35 rounds.

\subsection{Considerations on Metrics}
\label{sec:met}

As shown above, each evaluation work uses a particular scoring metric. Although accuracy is the most common metric, detection rate and F1-score are commonly employed as well. At any rate, this variety of metrics makes comparing solutions a complex and non-intuitive process. For this reason, we strongly believe that standardizing the evaluation process under a reliable metric is imperative. Assuming the intrusions as the positive class, the use of detection rate is not a fair practice due to a possible high amount of false positives. Thus, the best-practice is for it to be accompanied by the false positive rate (FPR). 

Among the metrics in the literature, F1-score is the most complete one due to its ability to wrap up precision and recall. However, as true negatives are not taken into account, F1-score could be problematic in asymmetric scenarios where the negative class is the minority (i.e., in a critical scenario where attacks are more common than normal traffic).

Although its utilization is not so popular, kappa statistic~\cite{kappa} is another interesting metric that quantifies the behavior of a predictive model in contrast to a random chance detector~\cite{borj196}. It works similarly to a correlation coefficient, rating model performance between $[-1,1]$. A value of 1 represents a complete agreement, 0 means no agreement or independence and, finally, a negative value implies that the predictive model is worse than random \cite{kappa2}. The adoption of the kappa statistic over the commonly used metrics is highly recommended due to its reliability and interpretation simplicity.

At any rate, we advocate the use of 2D graphical metrics contrasting standard statistical metrics such as positive predictive value (PPV), true positive rate (TPR) or false positive rate (FPR). Combining the ROC space (\textit{TPR vs FPR})~\cite{ROC} and the PR space (\textit{PPV vs TPR} )~\cite{prroc} with their corresponding curves and areas under the curves (AUCs) is a solid option to obtain a richer evaluation of most FL-ID models when compared with the metrics above. The graphical representation of those 2D metrics delivers complementary information. The ROC curve gives equal importance to positive and negative classes, whereas the PR curve is more informative in skewed scenarios, focusing on the positive class by penalizing false negatives considerably. Albeit to a lesser extent, the ROC curve penalizes a na\"{i}ve model behavior in unbalanced problems, when the majority of positive samples are predicted as negative. Therefore, it is equally capable of penalizing poor performance of predictive models in the minority (positive) class.

Figure~\ref{fig:roc_pr} shows how proposed graphical metrics are expected to behave in  some illustrative scenarios, as follows:
\begin{itemize}
    \item S1: absence of false negatives and absence of false positives (ideal case).
    \item S2: absence of false negatives and abundance of false positives.
    \item S3: abundance of false negatives and absence of false positives.
    \item S4: only false negatives and only false positives (worst case).
    \item Dashed line (TPR vs FPR): random guess.
    \item Dashed line (PPV vs TPR): variable baseline.
\end{itemize}

The penalization difference exposed in previous paragraphs is illustrated in Scenario \textit{S3}, as the PR curve shows a greater distance from the default baseline than the ROC curve from the random guess. That is especially noticeable when the positive is the extremely minority class. However, relying exclusively on the PR curve is not recommended due to its asymmetry -- not considering true negatives (Scenario \textit{S2}). Therefore, we believe that combining both graphical representations should be the best practice to make richer interpretations of the results, instead of other more commonly used metrics.

However, working with FLS entails the consideration of additional performance indicators. Measuring the variability of required communication rounds and elapsed time to achieve model convergence (i.e., until the learning model reaches a predefined quality threshold under a reliable metric), subject to a changeable number of parties (scalability), is important. In the same way, measuring the time required to perform a complete communication round in different bandwidth scarcity scenarios is interesting to evaluate the robustness and resilience of the system.

%%%%%%%%%%%%%%%%%%%%%%%%%%%%%%%%%%%%%%%%%%%%%%%%%%%%%%%%%%%%%%%%%%%%%%
\section{Future directions}
FL is a relatively new ML technique that is rapidly expanding due to its many benefits, as described in Section~\ref{sec:fl}. Thus, new variants and approaches are regularly being proposed, motivating the importance of this technology within the scientific community. Nevertheless, the default adoption of FL by most of the IDS is vanilla FedAVG. We foresee that IDS will benefit from improved FL developments and take advantage of variants that use this or other features of collaborative learning~\cite{lille}.

Furthermore, existing FL anomaly detection systems could be adapted to create and improve new FL-IDS. As mentioned in Section~\ref{lab:lstm}, technology that was not originally developed with ID proposes may present novel ideas that could be applied to FL-IDS. Hence, leveraging existing research, involving DAD, ID and FL could trigger the discovery of new advances. Currently, existing literature merging FL and IDS use simple IDS schemes based on anomaly detection and batch learning schemes. As shown in Section~\ref{sec:IDS}, incremental learning techniques and other detection architectures (misuse-based and hybrid ones) exist but have not been exploited yet to the context of FL-IDS.
Additionally, current FL-IDS do not unleash the full potential of the dedicated FL frameworks. Building FL models from scratch could be a time-consuming and ineffective task, considering the availability of advanced FL development libraries, e.g., TensorFlow federated, PySyft and FedML~\cite{fedml}.

As mentioned in Section \ref{sec:met}, the adoption of a standardized reliable metric is essential to perform fair evaluations of future FL-IDS models. We encourage the combined usage of both \textit{ROC} and \textit{PR} spaces as the default metric in order to provide a richer global picture in the different ID scenarios. Similarly, there is also a lack of standardization in terms of datasets as explained in Section \ref{sec:datasets}. Although dataset customization is an indicator of innovation, a series of constraints have to be considered in order to preserve quality standards. Particularly, we consider that providing raw \textit{pcap} files is important to give researchers the ability to fit data into their designed models.

Moreover, working with real-time data is not fully covered by current state-of-the-art approaches. Firstly, database updates by network traffic stream are not attended by default. Instead, the predominant trend is to evaluate models with generic preconceived datasets and extrapolate the learned models with no additional guarantees. Moreover, departing from a learned model, the detection of new threats is not trivial for current FL-IDS. A weight-based prototype should be introduced to help models to converge faster into new intrusion identification.

Finally, new generation (meta) FL-IDS should meet self security requirements. Avoiding model poisoning and malicious parameter update packets need to be part of security standards. In other words, FL-IDS should come with self-defense mechanisms. For instance, the incorporation of blockchain technology which can make the learning task more transparent and traceable~\cite{blockfla}.

\section*{Acknowledgments}
\noindent  This work is supported by the Basque Government (projects ELKARTEK21/89 and IT1244-19) and by the Spanish Ministry of Economy and Competitiveness MINECO (PID2019-104966GB-I00). Dr. Javier Navaridas is a Ram\'on y Cajal fellow from the Spanish Ministry of Science, Innovation and Universities (RYC2018-024829-I).

\bibliographystyle{IEEEtran}
\bibliography{fl-ids}

\end{document}